# The Entropy of Morbidity Trauma and Mortality

CE Neal-Sturgess, PhD.

Emeritus Professor of Mechanical Engineering

The University of Birmingham UK.

"..a living organism tends to approach the dangerous state of maximum entropy, which is death." Erwin Schrodinger, What is Life? 1944.

**Abstract**

In this paper it is shown that statistical mechanics in the form of thermodynamic entropy can be used as a measure of the severity of individual injuries (AIS), and that the correct way to account for multiple injuries is to sum the entropies. It is further shown that summing entropies according to the Planck-Boltzmann (P-B) definition of entropy is formally the same as ISS, which is why ISS works.

Approximate values of the probabilities of fatality are used to calculate the Gibb's entropy, which is more accurate than the P-B entropy far from equilibrium, and are shown to be again proportional to ISS. For the categorisation of injury using entropies it is necessary to consider the underlying entropy of the individuals morbidity to which is added the entropy of trauma, which then may result in death. Adding in the underlying entropy and summing entropies of all AIS3+ values gives a more extended scale than ISS, and so entropy is considered the preferred measure.

A small scale trial is conducted of these concepts using the APROSYS In-Depth Pedestrian database, and the differences between the measures are illustrated. It is shown that adopting an entropy approach to categorising injury severity highlights the position of the elderly, who have a reduced physiological reserve to resist further traumatic onslaught.

There are other informational entropy-like measures, here called i-entropy, which can also be used to classify injury severity, which are outlined. A large scale trial of these various entropy or i-entropy measures needs to be conducted to assess the usefulness of the measures. In the meantime, an age compensated ISS measure such as ASCOT or TRISS is recommended.

**Injury Severity Scaling**

Injury scaling, as a means of classifying the severity of impact trauma has a long history. Some of the earliest research into impact trauma was conducted at Cornell University Medical School in 1952 by De Haven and colleagues [1], and was related to aircraft crashes. The sixties saw many developments when a number of first generation methodologies were also proposed by:

Robertson et.al. [2], Nahum et.al. [3], Mackay [4], Van Kirk and Lange [5], States and States [6], Keggl [7], and Cambell [8].

In 1968 Ryan and Garrett [9] revised De Haven's scale, and considered energy dissipation, as well as threat to life, as criteria. The Comprehensive Research Injury Scale (CRIS) was developed using these concepts as shown in Table 1.

|         | Energy Dissipation | Threat to Life |
|---------|--------------------|----------------|
| Level 1 | Little             | None           |
| Level 2 | Minor              | Minor          |
| Level 3 | Moderate           | Moderate       |
| Level 4 | Major              | Severe         |
| Level 5 | Maximum            | Maximum        |

Table 1: Energy Dissipation and Threat to Life

The Abbreviated Injury Scale (AIS) was first officially published in 1972, revised in 1974 and 75 and published in manual format in 1976 [10]. Revisions were published in 1980, 85, 90, 98 and 2005 [11]. AIS is usually described as a non-linear ordinal scale, this is incorrect. If the genesis of AIS is followed it is obvious that it is a non-linear integer scale related to energy dissipation. It is integer simply because no fractional AIS measures have been introduced, as the clinical resolution would not support such fine scale measures. The AIS score has proven to be the "system of choice" [1], and has been documented in many articles [12, 13]. A number of user groups have modified the basic AIS scale to account for particular types of injury and harm, and a case has been made for unification[14].

Baker et.al. [15, 16] studied over 2000 vehicle occupants, pedestrians, and other road users, and found that the AIS score was a non-linear predictor of mortality. This is not a fundamental problem as non-linearities can be easily accommodated. However, it was found that the death-rate of persons with two or more injuries was not simply the sum of the AIS scores. This led to the introduction of the empirical Injury Severity Score (ISS) as a means of linearising the data with regard to the probability of fatality. The ISS is the sum of squares of the maximum AIS code (MAIS) in each of the three most severely injured body regions [15].

ISS has been criticized as it only allows the counting of one MAIS value per body region, where in reality there are often more that on MAIS value in a given body region.  The New Injury Severity Scale (NISS) was introduced [13] to allow the counting of more than one MAIS value for a given body region. It was shown that NISS is marginally more effective than ISS. Both ISS and NISS are calculated on the basis of ordered triplets, and so there are a significant number of ISS or NISS values that cannot be achieved in practice.

**Thermodynamic Entropy as a measure of Trauma**

**Single Injuries**

Biomechanical injuries are the result of the separation (fracture, shearing, tearing or rupture) of biological tissues. In a living biological system tissue separation is often followed by repair, however this does not mean that the process of separation is "reversible" in a thermodynamic sense. It is therefore considered that injuries in Impact Trauma may be viewed as "mechanical dissipative processes" i.e. they require an expenditure of work, and that they should be consistent with the Second Law of Thermodynamics and describable in terms of Irreversible Thermodynamics [17]  Irreversible thermodynamics features the Clausius-Duhem Inequality (CDI), which gives a formalism for deriving constitutive equations for irreversible dissipative processes.  This has been applied to a number of flow and damage problems in continuum mechanics since the 1980's  [18-20].

Using the CDI Sturgess [21] showed that during a crash pulse the injury severity, as measured by AIS, is proportional to the maximum rate of entropy production, expressed as the Peak Virtual Power (PVP).  This criterion results in the following set of equations:

$$PVP \propto \hat{a}V \propto \bar{a}^2 \Delta t \propto VC_{max} \propto V^3$$
(1)

Where:

    PVP     = Peak Virtual Power

    $\hat{a}$     = maximum acceleration

    $\bar{a}$     = mean acceleration

    V     = velocity

    $\Delta t$     = time interval

    $C_{max}$     = maximum compression

The criteria are, in order from left to right, the Margulies and Thibault criterion [22], the Head Impact Power (HIP) [23, 24], and the Viscous Criterion  [25].  Therefore PVP is an overarching criterion, which encompasses all the other scientifically valid injury criteria, and shows that the maximum rate of entropy production is a valid means of assessing injury severity.

In a simplified Newtonian impact the PVP is proportional to the change in velocity (Delta V) cubed for all body regions, as shown in Fig. 1. The correlation coefficients ($R^2$) are between 0.8 and 0.98 dependent on body region.

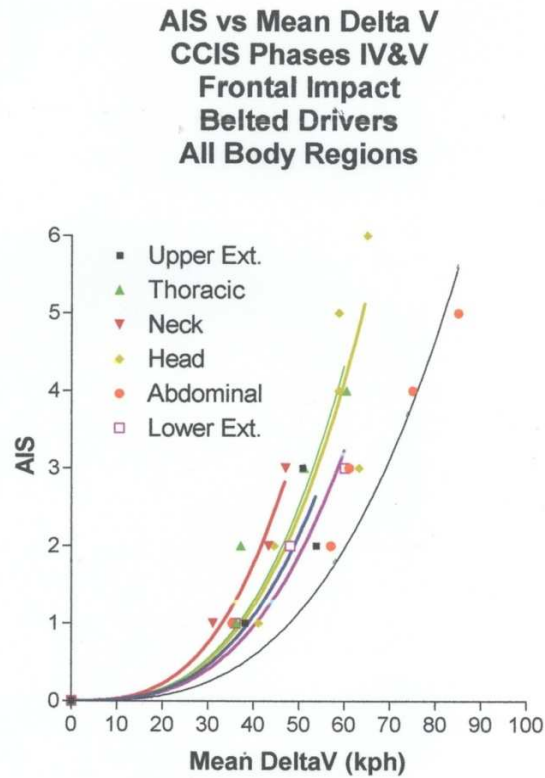

Figure 1:  AIS versus DeltaV

A way of visualising this is that the injuries in a crash are proportional to the total energy transferred or total entropy, which are only known after the event.  During the crash pulse a suitable metric that is proportional to the total entropy is the rate of entropy production, and the total entropy is simply the integral of the rate.

There are three measures currently called "entropy" in current use.  The first is thermodynamic entropy invented by Clausius in 1862 [26], and the more recent related measure called informational entropy by Shannon [27]; here called i-entropy.  There is a further version of entropy used in the Table of Life [28], but that is not related to the other two.   Thermodynamic and informational entropy are linked by mathematics, but not by physics, as will be explained later.  In this paper entropy refers to thermodynamic entropy as originally intended by Planck, Boltzmann and Gibb's [29], and i-entropy is used to refer to informational entropy.

Entropy is an abstract quantity that has caused considerable confusion since its invention by Rudolph Clausius in 1862 [26], the mathematics are often difficult, but in principle it is a simple concept. If energy is conserved according to the first law of thermodynamics, and if the Gibb's free-energy of a system (the ability to do work) actually does work, then the free-energy decreases [30]. However, if energy is conserved then "something" must increase to compensate for the decrease in free-energy, that something is entropy. Entropy is a composite word 'en' from energy, and 'tropy' from the Greek 'trope' meaning form. Hence entropy means the energy of the change of form, and it is the irreversible component of energy transfer that changes the form of the system or body. It is intuitively obvious that it can be connected to injury, which is demonstrably an irreversible change in form.

Entropy is often described as the tendency for the descent into chaos and disorder. However, a more rigorous view is that it is the tendency to "thermal equilibrium", which is the state of maximum entropy. If a system is in thermal equilibrium then all constituent parts are at the same energy level, hence no energy transfer can take place, and no work done.

**Multiple Injuries**

If injury severity is proportional to the total entropy in the process, then it should be describable in terms the statistical mechanics of Planck-Boltzmann [30]. Statistical mechanics is the branch of mechanics concerned with relating the microscopic properties of a body to the macroscopic observable properties of that body, using probability theory based on physical processes. This is in contrast to statistics which is the branch of mathematics which deals with the collection and analysis of data, without necessarily any recourse to physical principles.

The form of entropy described by Planck-Boltzmann [29] is that of a logarithmic form:

$$S = -k \sum lg\, p_i \qquad (1)$$

Where:

$S$ = entropy

$k$ = Boltzmann's constant

$p_i$ = the probability of the state 'i', interpreted here as proportional to AIS

This again has caused much confusion. The reasoning behind this description is that Boltzmann wanted to describe a combined state of two independent probabilities say $p_1$ and $p_2$ co-existing, for which the combination is described by the product law of probabilities [31]. However, entropy is an energy term (JK$^{-1}$ a scalar) and hence additive, so to express the product probability law as a summative process Boltzmann deduced that entropy must be represented in a logarithmic form such that:

$$P = p_1 . p_2 \propto \sum_{i=1}^{i=i} lg\, p_i \qquad (2)$$

This is the basis of the famous expression for entropy inscribed on Boltzmann's tomb, equation 1, which was first described by Planck [32].

Gibb's [29] used a more mathematically rigorous derivation of entropy to give:

$$S = -k \sum p_i \lg p_i \qquad (3)$$

Which is the usual form of entropy that is now used, as it remains valid far from equilibrium [29]. The base of the logarithm used is not relevant, it can be either 2 or 10 without loss of generality.

The statistical mechanics interpretation of entropy by Boltzmann was to take a high level viewpoint, as to compute the observables from Newtonian mechanics for each molecule in a gas is simply impossible; this is similar to the mechanics of Trauma. To know all the interactions going on in creating an injury is impossible, so a high level descriptor like Entropy, which is based on statistical mechanics and sound physical principles, is a prime candidate.

To utilise the principles of statistical mechanics to injury, the problem now arises as to the interpretation of the probabilities in equation 3. One of the outstanding problems in scaling injury severity is how to combine different injuries to estimate fatality. So the question is can a measure of entropy be defined which represents morbidity, with a limit of fatality (mortality), and if so what do the probabilities represent?

If the probability of fatality is plotted for different values of MAIS, then figure 2. results:

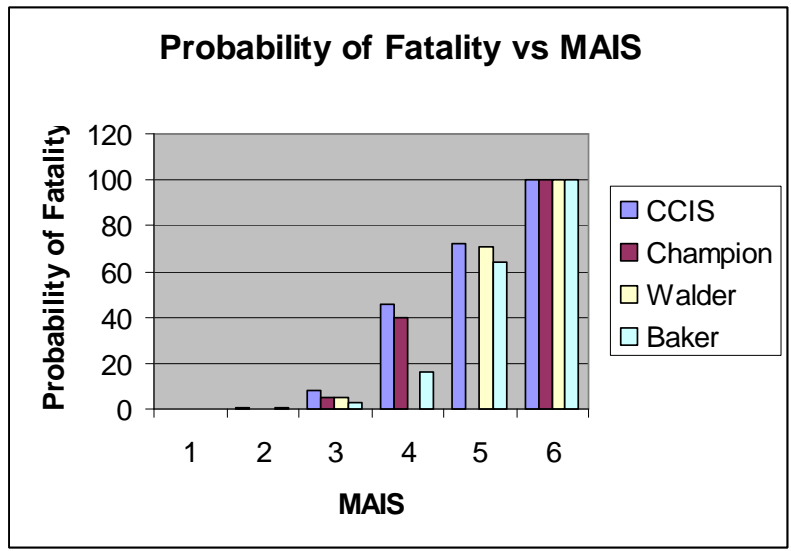

Figure 2. Probability of Fatality versus MAIS

The data from the various studies are [16, 33, 34]  From this figure it can be seen that as the value of the MAIS increases then there is a non-linear relationship with the probability of fatality.  The problem with this representation, as noted by Baker et.al. [16], is that a MAIS5 (for instance) can include any number of other AIS5's, and also any number of other lower AIS values.  This led to the formulation of the empirical Injury Severity Score ISS, which is the sum of squares of the three MAIS values for different body regions, which gives a reasonable predictor of fatality as shown by Baker et.al's data in Fig 3 below.  However, to date there has not been any reasons advanced as to why this procedure works, which is addressed here.  Figure 3. has been redrawn form Baker et.al's original Fig 4 [16], with the addition of the weighted aggregate results labelled 'All'.

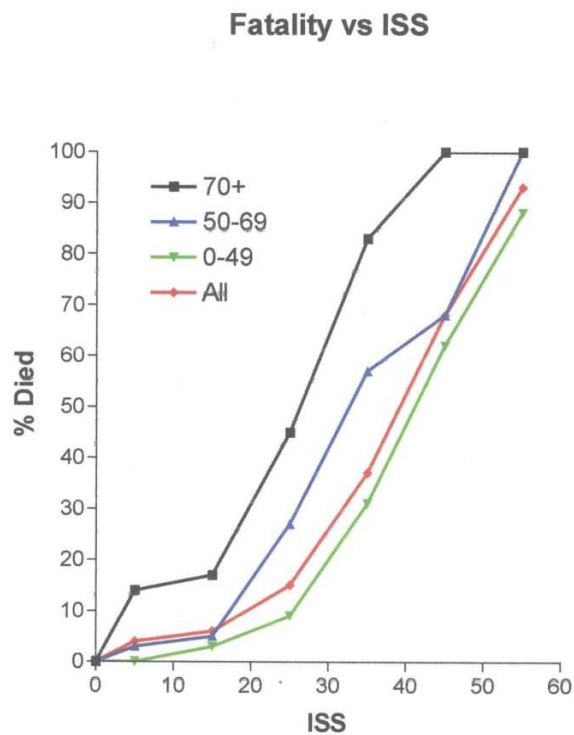

Figure 3.  Probability of Fatality versus ISS from Baker et.al.

It is also worth noting from fig 3. that above an ISS value of around 20 for the aggregate curve (All), then the curve is virtually linear, which was the feature Baker et.al sought.

From Fig.3 it can be seen that there is a very large difference between the probabilities of fatality for the 70+ group and the other age groups, particularly so for low values of ISS.  It is also obvious that aggregating the data largely ignores the effects of the 70+ due to the weighting factors. Baker et.al were the first to draw attention to the phenomena, and although it is often mentioned in automotive safety studies [35] little has yet been done about it.  It is becoming of ever greater significance as the demographics are changing to give a much older population.

A problem with ISS is that the score is based on the MAIS values for the three most severely injured body regions, however if it frequently found that one body region may have more than one MAIS injury, so the New Injury Severity Score (NISS) was devised.  The NISS score takes the three highest MAIS values irrespective of body region, and has been used with varying degrees of success [36, 37].  NISS is also proportional to Entropy.

The fundamental problem with ISS, as recognised by Boyd et.al. [38], is that ISS essentially only measures the traumatic insult, and the combined effect of the traumatic insult and the persons underlying medical and physiological reserve is also very important.  This led to the ASCOT scoring system, and Champion has shown by using logistic regression that ASCOT has a better predictive capabilities than ISS or NISS [39]. This research finally led to the TRISS method (Trauma and Injury Severity Score), which includes the ISS score, the RTS (Revised Trauma Score, and the patient's age) [38], which is available as an online calculator [40]. Age is a very important variable in trauma, and Both ASCOT and TRISS are very coarse grained with respect to age, leading to mis-diagnosis in a number of seriously injured casualties [41].

A considerable amount of work has been done on the subject of entropy and ageing, beginning with Schrodinger in 1944 [42], and followed by Strehler [43, 44].  It is a reasonable assumption that entropy increases throughout life, either through genetic mutations , disease , or miss-repair [45-48], probably in a non-linear manner.  This is shown schematically in Fig.4.

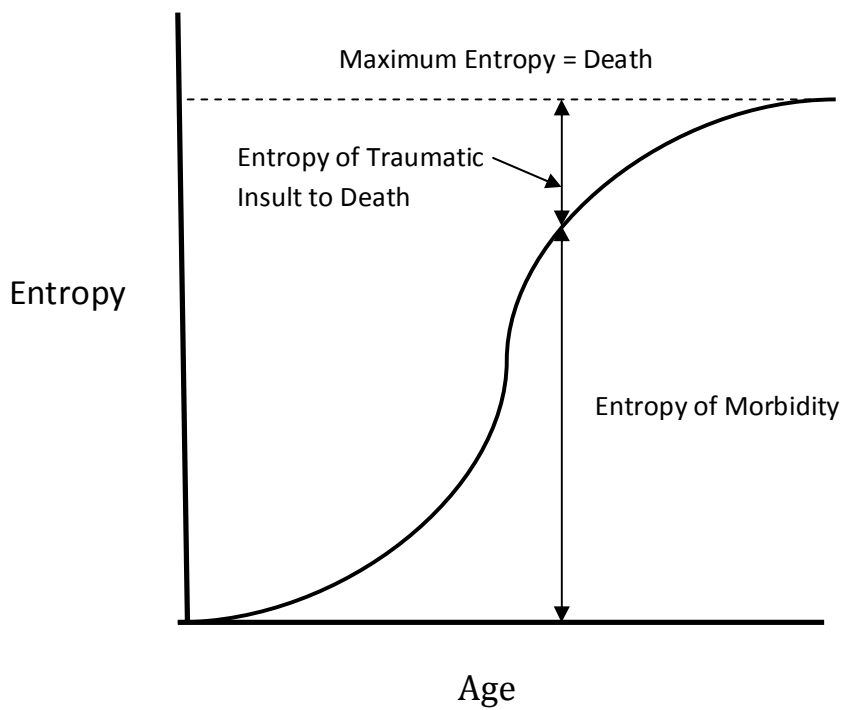

Fig. 4. Schematic of the Entropy of Life Curve.

A traumatic insult only needs to add the entropy between the entropy of morbidity, and the maximum entropy that life can withstand to induce death, called the entropy of mortality $S_{Mt}$, as shown in Fig.4. This model successfully accounts for the observations that the "elderly" die at much lower injury severities than younger persons, as characterised by the entropy of morbidity, which includes pre-existing medical conditions. If this function could be accurately determined it could be used as real (entropic) age as opposed to calendar age.

Life expectancy (LE) has been rising for a considerable time, however the Disability Free life Expectancy (BFLE) has been rising at a slower rate [49, 50]. This has led to a larger proportion of the population living with accumulating illness and disability. It is this elderly part of the population which is increasing and leading to increased mortality at low injury levels [47, 51].

As entropy is additive, then denoting the entropy of mortality as $S_{Mt}$, the accumulated entropy of morbidity at a given age as $S_{Mb}$, and the entropy of the traumatic insult as $S_T$ then:

$$S_{Mt} = S_{Mb} + S_T \tag{4}$$

Then the Gibb's entropy of trauma (equation 3), can be restated as:

$$S_T = k \sum p_i \lg p_i \tag{5}$$

And so the full expression is:

$$S_{Mt} = S_{Mb} + k \sum p_i \lg p_i \tag{6}$$

The values for $S_{Mb}$ and the constant k will be strong functions of age, as explored later.

If the process of determining the ISS value is examined, then it is:

$$ISS = MAIS_1^2 + MAIS_2^2 + MAIS_3^2 \tag{7}$$

Where the subscripts refer to the three different body regions counted. If logarithms are taken for the special case where the MAIS values for the different body regions are the same, then equation 7 becomes:

$$\lg ISS = 2 \sum_{i=0}^{i=3} \lg (MAIS_i) \tag{8}$$

Which is of the same mathematical form as equation 1. This equation is plotted on logarithmic axes in Fig.5, where it can be clearly seen to give a proportional relationship between an entropic measure and ISS.

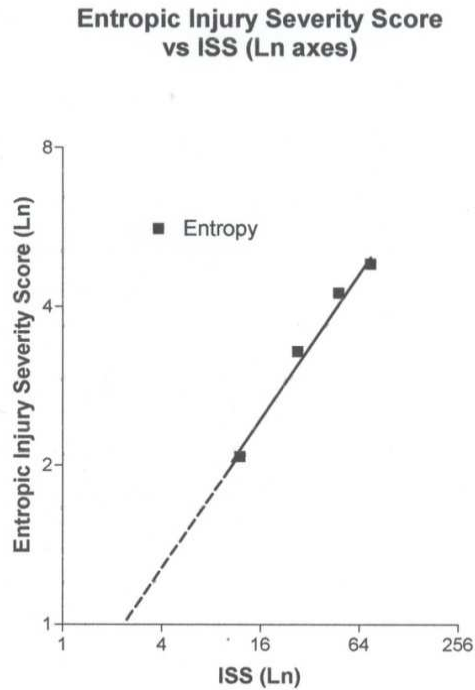

Figure 5.  Entropy (equation 8) versus ISS

That entropy is proportional to ISS is shown by the non-zero intercept, and a slope which is not unity.  It can therefore be seen that what is being calculated in ISS is mathematically similar to the logarithmic sum of probabilities.  It is summing entropy, which is considered to be the reason why ISS works.

From Fig 3. it can be seen that for the lower range of values for ISS, i.e. take 25 for instance, then this represents either one AIS5, or one AIS4 plus one AIS3.  Similarly for ISS 16, this can only represents one AIS4.  ISS 9 can only represent one AIS3, and ISS 4 can only be one AIS2.  This highlights one of the problems with ISS indicated by Baker et.al.[16], that as ISS is calculated from a set of ordered triplets. then there are a significant number of ISS values which cannot be achieved in practice.

In the initial evaluation of ISS it was decided that an ISS value of 75 corresponds to a very high probability of fatality, and was chosen as the cut off point.  This obviously corresponds to three MAIS5's, or AIS5's if they are the only injuries.

Utilising the data from Baker et.al. (Fig 3.) the probabilities are as shown in Table 2.

| ISS | MAIS/AIS | % Probability of Fatality of AIS (Elderly) | Entropy $S_T$ Aggregate (Elderly) |
|---|---|---|---|
| 0 | 0 | 0     (?)  | < 0.0   (0) |
| 1 | 1 | <1    (?)  | < 0.0   (0) |
| 4 | 2 | 1     (14) | < 0.0   (16) |
| 9 | 3 | 3     (15) | 1.43    (18) |
| 16 | 4 | 6    (17) | 4.7     (20) |
| 25 | 5 | 15   (50) | 18      (84) |

Table 2: Measures of Injury Severity

It is obvious from Table 2 that the elderly have a much higher probability of mortality and entropy than the under 70's. Taking the same cut off point, and choosing an entropy of 100 to correspond to death, ignoring the minus sign as this is only convention, then equation 5 becomes:

$$100 = k(3)(15)(lg15) \tag{9}$$

therefore k = 2, and equation 5 becomes:

$$S_T = 2 \sum_{i=0}^{i=i} p_i lg p_i \tag{10}$$

For ISS this was equated to the empirical function describing the proportion of those who died. It is obvious comparing equations 8 and 10 that the two are proportional, as will be seen later. Note that as the entropy of death has been equated to 100, and entropy is additive, then intermediate values can be considered a percentages.

For a test bed for the correlation between entropy and ISS, the APROSYS In Depth Pedestrian Database was utilised [51, 52]. This database is very detailed, and was compiled to enable sufficient in-depth data to be incorporated so that the crashes could be simulated to a high degree of accuracy. The database is not truly representative only including 70 cases, and biased towards elderly fatal injuries. The database is nonetheless a real world database, and suitable for the trial. In the APROSYS In Depth Database 50% (7/14) of the elderly fatally injured were at AIS 3, and in general the elderly need to be considered as a separate category, as done later.

The results are shown in Fig.6. What is done in Fig.6. is that the entropies of all AIS3+ injuries are summed in accordance with the probabilities from Table 2.

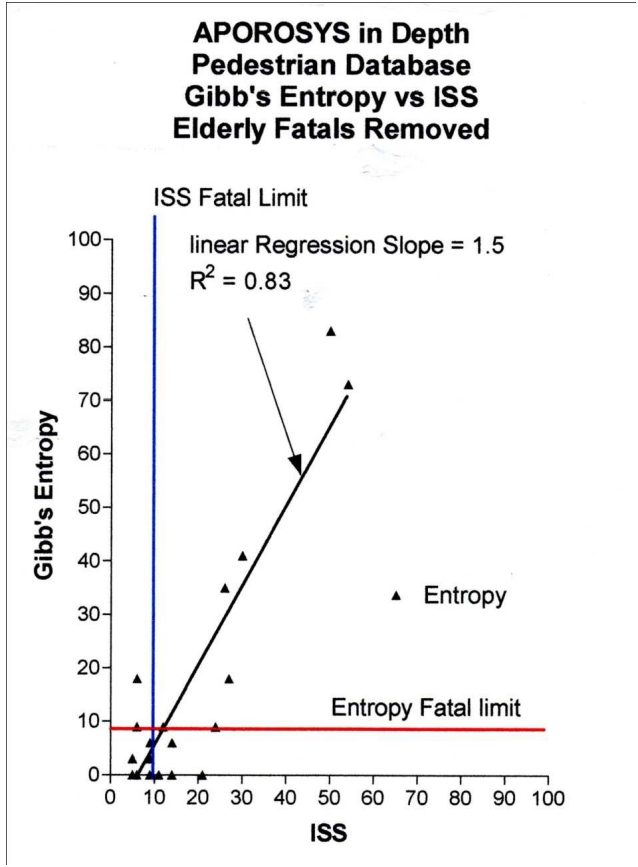

Fig.6. Gibbs Entropy vs ISS

The actual value of the % probability has been used, rather than a fractional term, as the logarithmic function is not well behaved between 0 and 1, and the constant chosen corrects for this.

As can be seen from Fig.6. that the relationship between entropy and ISS is proportional, as would be expected from the above, but they not the same. In particular there is a considerable non-zero intercept (representing $S_{LM}$, and the slope of 0.8 is significantly less that unity. This shows that for a given spectrum of injuries summing probabilities at all injuries of AIS > 3 as entropy, gives a more extended measure than ISS. This would aid differentiation between categories of injury, and is considered a desirable attribute.

When epidemiologically studies are done the researchers often choose different values of AIS to represent "serious" injury. From Table 2, and Fig .6. it is obvious that the contribution to entropy for any AIS < 3 is negligible, therefore serious injury should be defined as AIS3+. This is consistent with TRISS and ASCOT.

It is difficult to estimate values for $S_L$ as it obviously depends on the medical and physiological condition of each person. However, it is considered that approximate values can be obtained by considering the upper and lower limits of the entropies for each age group. For the middle age group the highest non-fatal entropy of trauma was 73, and the lowest fatal entropy of trauma 22. Subtracting these values from 100 and taking the average gives the $27 < S_{LM} < 78$ with the average $S_{LM}$ ($\bar{S}_{LM}$) = 53, assuming that the distribution is normal. This needs investigating as to what kind of distribution it is, and to set confidence limits. This is surprisingly high, and indicates that there are probably many middle aged people with significantly compromised medical and physiological reserve.

[53]Calculating the constant k in equation 6 for the limit of three AIS5's gives k = 0.9, and so the full entropy equation for the middle aged is:

$$S_{MM} = 53 + 0.9(ISS) \qquad (11)$$

This function is plotted in Figure 7.

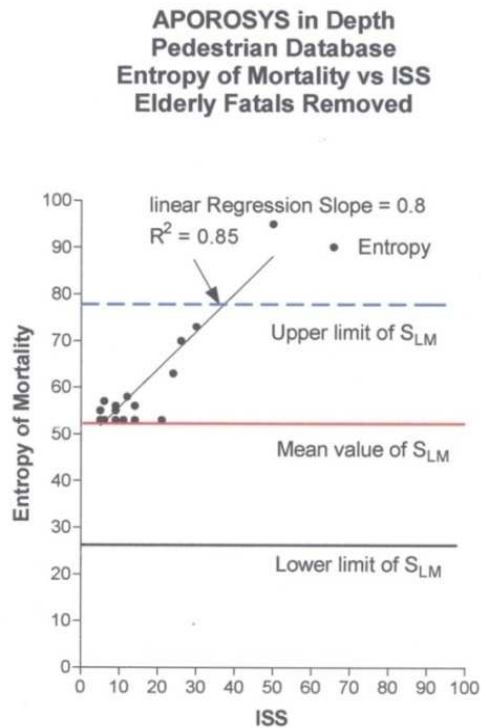

Figure 7. Entropy (equation 11) versus ISS

Conducting the same calculations for the elderly age group gives the highest elderly non-fatal entropy = 47, and the lowest elderly fatal entropy = 3. Therefore for the elderly $57 < S_{LE} < 97$, with $\bar{S}_{LE}$ = 72, and $k_E$ based on 3 AIS3's as 0.5, as AIS3 appears to be the relevant level for the elderly. The equation is:

$$S_{ME} = 75 + 0.5(ISS) \qquad (12)$$

It should be noticed that with the upper value of $S_{LE} = 97$, then any person in this condition would have essentially no physiological or medical reserve left, and are likely to be fatally injured in a collision of any severity. Further, from Table 2 with a residual $S_{LE} = 75$ then the addition of one AIS5 injury would give a value of $S_{DE} = 117$ which is not survivable, and actually represents overkill, as shown in Fig.8.

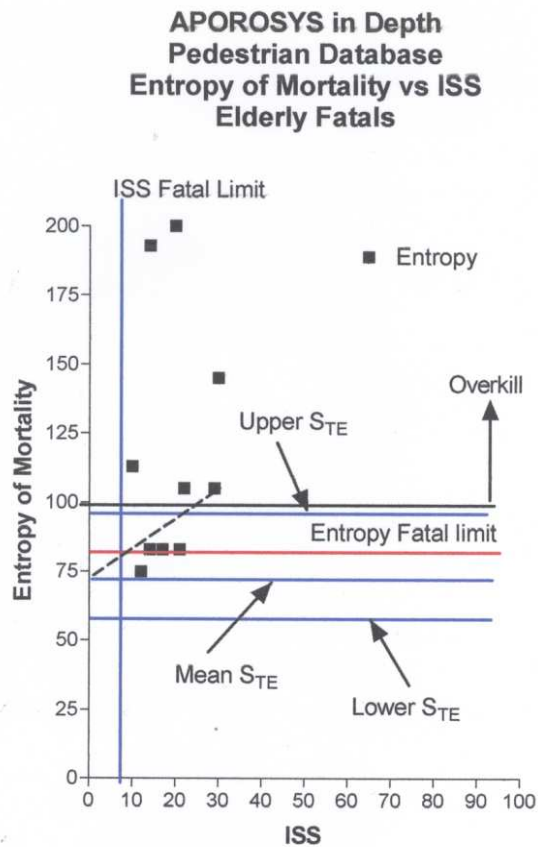

Fig.8: Entropy of Mortality vs ISS (equation 12) for the Elderly

It should also be noted that many researchers consider the Pre-Existing medical Conditions (PEMS's) are responsible for the reduced ability to survive of the elderly and medically compromised individuals, and the intercept of 75 at ISS = 0 includes PEMC's [54]. This approach should give results similar to ASCOT and TRISS, but it is both physically based and easier to apply.

**Informational Entropy (i-entropy) as a measure of Trauma**

In the earlier sections of this paper only the effects of thermodynamic entropy were considered, and successfully accounted for the reason ISS works. There is however another measure known as informational entropy derived by Shannon in1948 [27] . Shannon's entropy, here denoted i-entropy, is a measure of the uncertainty of a random variable, and results from taking a logarithmic sum of probabilities which results in the same form of equation as the of Gibb's entropy. It has been called entropy without having the linkage to energy transfer, which is the basis of thermodynamic entropy. This is somewhat unfortunate as Fisher [55] ] had also defined a logarithmic sum of probabilities in 1932 to use in the statistics of meta-analysis, but he did not call it entropy. Therefore entropy and i-entropy notionally at least refer to different entities. However if Lloyd's view [32] that the universe only consists of information, and the change of form is due to irreversible bit-flips, which require energy, then the two concepts of entropy and i-entropy can be reconciled.

There have apparently been two attempts to use i-entropy in trauma studies. The first is a series of papers on using the i-entropy of heart rate monitor signals to predict the patient's well being [56]. A statistical measure used in the Table of Life called entropy is very different from the entropy measures used here [28] . However, a further entropy-like measure that could be used in injury severity scaling is the statistical term "conditional entropy" [53]. This again has the same form as Gibb's entropy (equation 3) and describes the intersection of two sets $A \cap B$ as shown in fig.9.

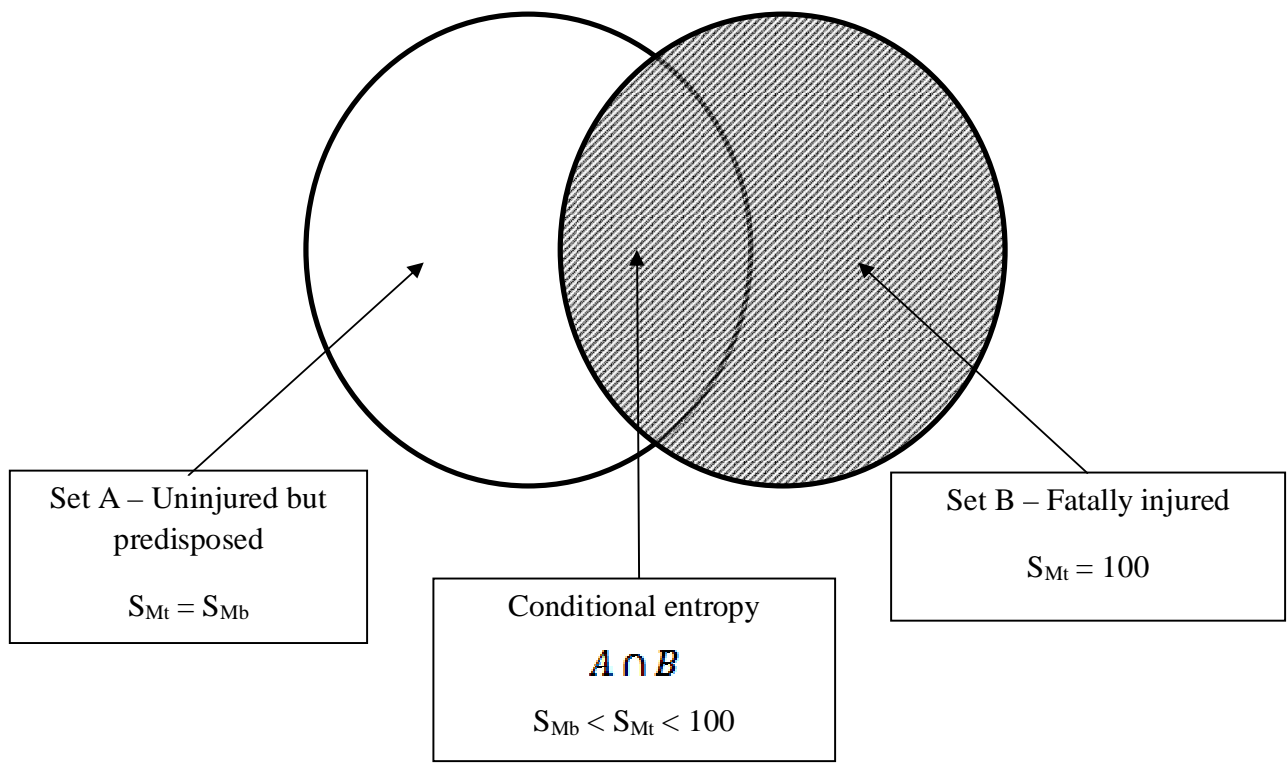

Figure 9: The intersection of two sets defined by the conditional entropy.

This gives a readily understandable picture of the injury problem, and is a useful way of depicting the meaning of the entropy of Trauma. Everyone enters a traumatic episode with predisposing factors represented by the entropy of morbidity ($S_{Mb}$), hence there is always an intersection of the sets A∩B which is added to by the entropy of trauma. Hence the two systems of either statistical mechanics (thermodynamic entropy) or statistics (i-entropy) in this instance lead mathematically to the same conclusions.

A large scale trial of these various entropy or i-entropy measures needs to be conducted to assess the usefulness of the measures. As summing probabilities or summing squares of AIS (the two are in principle proportional) of all AIS3+ injuries as entropy gives a more extended scale than ISS, it is considered the preferred measure. In the meantime ASCOT or TRISS, if they can be applied, appear to give advantages over NISS or ISS.

**Conclusions**

- It has been shown that thermodynamic entropy can be used as a measure of the severity of individual injuries as measured by the Abbreviated In jury Scale (AIS)
- It is shown that the correct way to account for multiple injuries is to sum the entropies, and it is further shown that summing Planck-Boltzmann entropies is in principle the same as ISS
- An entropy measure of the severity of injury requires that the entropy accumulated in life prior to the injury is needed to add to the entropy of trauma to predict mortality.
- As summing probabilities of all AIS3+ injuries as entropies gives a more extended scale than ISS, and is considered the preferred measure.
- There are other informational entropy-like measures, called i-entropy, which can also be used to classify injury severity, and are proportional to thermodynamic entropy.
- A large scale trial of these various entropy and/or i-entropy measures needs to be conducted to assess the usefulness of the measures. In the meantime ASCOT or TRISS appear to give advantages over ISS or NISS.

**References:**

1. Petrucelli E (1993) Injury Scaling Systems and Disability Measurements  AAAM??
2. Robertson JS, McLean AJ and Ryan GA (1966) Traffic Accidents in Adelaide South Australia   Australian Road Research Board
3. Nahum A, Siegel AW and Hight PV (1967) Injuries to Rear Seat Occupants in Automobile Collisions  11th STAPP Car Crash Conference
4. Mackay GM (1968) Injury and Collision Severity  12th STAPP Car Crash Conference
5. Van Kirk DJ and Lange WA (1968) A Detailed Injury Scale for Accident Investigation  12th STAPP Car Crash Conference

6.  States J and States D (1968) The Pathology and Pathogenesis of Injuries Caused by Lateral Impact Accidents   12th STAPP Car Crash Conference
7.  Keggl K (1969) Yale Trauma Scale    University of Yale Medical Centre, New Haven NY
8.  Cambell EOF (1969) Collision Tissue Damage Record    Traffic Injury Research Foundation of Canada
9.  Ryan GA and Garrett JW (1968) A Quantitative Scale of Impact Injury    Cornell Aeronautical Laboratory, Buffalo NY
10. Scaling CoI (1976) The Abbreviated Injury Scale 1976 Revision    AAAM
11. Gennarelli TA and Wodzin E (2006) AIS 2005: A contemporary injury scale  Injury 37, 1083-1091 DOI
12. Medicine AAfA (1983) Injury Scaling Bibliography    AAAM
13. Stevenson M, Segui-Gomez M, Lescohier I, Di Scala L and McDonald-Smith G (2001) An overview of the injury severity score and the new injury severity score. Injury Prevention 7: 10-13 DOI 10.1136/ip.7.1.10
14. Garthe E and Mango N (1998) AIS Unification: The case for a Unified Injury System for global use  16th ESV Conference  NHTSA
15. Baker SPea (1974) The Injury Severity Score: Development and Potential Usefulness  18th AAAM Conference
16. Baker S and O'Neil B (1974) The Injury Severity Score:  A method for describing patients with multiple injuries and evaluating emergency care. Trauma 14: 183-196 DOI
17. Jou D, Casa-Vazquez J and Lebon G (1993) Extended Irreversible Thermodynamics  Springer-Verlag, Berlin, Heidelberg, New York
18. Krajcinovic D and Lamaitre J (1986) Continuum Damage Mechanics - Theory and Applications    Springer Verlag, Berlin
19. Lamaitre J and Chaboche J-L (1990) Mechanics of Solid Materials Cambridge University Press
20. Murakami S and Kamiya K (1997) Constitutive and damage evolution equations of elastic-brittle materials based on irreversible thermodynamics. Int J of Mechanical Sciences 39: 473-486 DOI
21. Sturgess CEN (2002) A Thermomechanical  Theory of Impact Trauma. Proc IMechE, Part D:  J of Automobile Div 216: 883-895 DOI
22. Margulies SS and Thibault LE (1992) A Proposed Tolerance Criteria for Diffuse Axonal Injury. J of Biomechanics 25: 917-923 DOI
23. Newman J, Barr C, Beusenberg M, Fournier E, Shewchenko N, Welbourne E and Withnall C (1999) A New Biomechanical Assessment of Mild Traumatic Brain Injury, Part 1 - Methodology  IRCOBI International Council on Biokinetics of Impacts.
24. Newman J, Barr C, Beusenberg M, Fournier E, Shewchenko N, Welbourne E and Withnall C (2000) A New Biomechanical assessment of Mild Traumatic Brain Injury, Part 2 - Results and Conclusions IRCOBI 2000 International Council on the Biokinetics of Impacts
25. Viano DC and Lau IV (1987) A Viscous Tolerance Criterion for Soft Tissue Injury Assessment. J of Biomechanics
26. Cardwell DSL (1971) From Watt to Clausius: The Rise of Thermodynamics in the Early Industrial Age. Heineman, London

27. SHANNON CE (1948) A Mathematical Theory of Communication. The Bell System Technical Journal 27: 379–423 DOI
28. Goldman N and Lord G (1986) A new look at Entropy and the Life Table. Demography 23
29. Jaynes ET (1965) Gibbs vs Boltzmann entropies. American Journal of Physics, 33: 391-8 DOI
30. Kondepudi D and Prigogine I (1998) Modern Thermodynamics John Wiley & Sons Ltd
31. Olofsson P (2005) Probability, Statistics, and Stochastic Processes Wiley-Interscience
32. Lloyd S (2008) The Digital Universe  Physics World
33. Copes WS, Sacco WJ, Champion HR and Bain LW (2007) Progress in Characterising Anatomic Injury 33rd Annual Meeting of the Association for the Advancement of Automotive Medicine, Baltimore, MA, USA, pp 205-218 DOI
34. Walder AD, Yeoman PM and Turnbull A (1995) The abbreviated injury scale as a predictor of outcome of severe head injury. Intensive Care Medicine 21: 606-609 DOI
35. Morris A, Welsh R, Frampton R, Charlton J and Fildes B (2002) An Overview of Requirements for the Crash Protection of Older Drivers   AAAM
36. BALOGH ZJ, VARGA E, TOMKA J, SÜVEGES G, TOTH L and SIMONKA JA (2003) The new injury severity score is a better predictor of extended hospitalization and intensive care unit admission than the injury severity score in patients with multiple orthopaedic injuries. Journal of Orthopaedic Trauma   17: 508-12 DOI
37. Honarmand A and Safavi M (2006) The New Injury Severity Score. Indian J Crit Care Med 10
38. Boyd CR, Tolson MA and Copes WS (1987) Evaluating Trauma Care: The TRISS Method. Trauma 27: 370-378 DOI
39. Champion HR, Copes WS, Sacco WJ, Frey CF, Holcroft JW and Hoyt DB (1996) Improved predictions from a severity characterization of trauma (ASCOT) over Trauma and Injury Severity Score (TRISS): results of an independent evaluation. Trauma 40: 42-49 DOI
40. Boyd CR, Tolson MA and Copes WS (1987) TRISS Calculator
41. Demetriades D, Chan LS, Velmahos G, Berne TV, Cornwell EE, Belzberg H, Asensio JA, Murray J, Berne J and Shoemaker W (1998) TRISS methodology in trauma: the need for alternatives. Br J Surg 85: 379-84 DOI
42. Schrodinger E (1944) What is life? Cambridge University Press  2007
43. Strehler BL (1960) Fluctuating energy demands as determinants of the death process (A parsimonious theory of the Gompertz function) In: al Se (ed) The Biology of Aging Amer. Inst. Biol. Sci., Washington, pp 309-314 DOI
44. Strehler BL (1962) Time, Cells, and Aging Academic Press, New York
45. Atlan H (1975) Strehler's Theory of Mortality and the Second Principle of Thermodynamics
    NASA-TK-X-74389
46. Riggs JE (1993) Aging, genomic entropy and carcinogenesis: implications derived from longitudinal age-specific colon cancer mortality rate dynamics. Mech Ageing Dev 72
47. Silva C and Annamalai K (2008) Entropy Generation and Human Aging: Lifespan Entropy and Effect of Physical Activity Level. Entropy 10: 100-123 DOI 10.3390/entropy-e10020100

48. Salminen A and Kaarniranta K (2010) Genetics vs. entropy: Longevity factors suppress the NF-kB-driven entropic aging process. Ageing Research Reviews 9: 298-314 DOI
49. (2006) Healthy Life Expectancy    Parliamentary Office of Science and Technology
50. Robine J-M, Mathers C and Brouard N (1996) Trends and Differentials in Disability Free Life Expectancy In: Caselli  GaL, A.D. (ed)  Health and Mortality Among Elderly Populations  Clarendon Press, Oxford
51. Carter EL, Neal-Sturgess CE and Hardy RN (2009) APROSYS in-depth database of serious pedestrian and cyclist road traumas. International Journal of Crashworthiness (APROSYS special issue)
52. Neal-Sturgess CE, Carter E, Hardy R, Cuerden C, Guerra L and Yang J (2007) APROSYS European In-Depth Pedestrian Database  ESV 2007
53. Korn TM and Korn GA (2000) Mathematical Handbook for Scientists and Engineers  Dover Publications, New York
54. McMahon DJ, Schwab CW and Kauder D (1996) Comorbidity and the Elderly Trauma Patient. World J Surg 20: 1113-1120 DOI
55. Fisher RA (1948) Combining independent tests of significance. American Statistician 2
56. Costa MD, Peng CK and Goldberger AL (2007) Multiscale Analysis of Heart Rate Dynamics: Entropy and Time Irreversibility Measures. Cardiovasc Eng